# Control of Neuronal Survival and Development Using Conductive Diamond

Samira Falahatdoost [1], Yair D. J. Prawer [2], Danli Peng [1], Andre Chambers [1,3], Hualin Zhan [1,4], Leon Pope [5], Alastair Stacey [6], Arman Ahnood [5], Hassan N. Al Hashem [5], Sorel E. De León [5], David J. Garrett [5], Kate Fox [5], Michael B. Clark [7], Michael R. Ibbotson [8], Steven Prawer [1], and Wei Tong [1*]

[1] School of Physics, The University of Melbourne, Parkville, Victoria 3010, Australia
[2] Department of Anatomy and Physiology, The University of Melbourne, Parkville, Victoria 3010, Australia
[3] Department of Mechanical Engineering, The University of Melbourne, Parkville, Victoria 3010, Australia
[4] School of Engineering, The Australian National University, Canberra, Australian Capital Territory 2601, Australia
[5] School of Engineering, STEM College, The RMIT University, Melbourne, Victoria 3000, Australia
[6] School of Science, STEM College, The RMIT University, Melbourne, Victoria 3000, Australia
[7] Department of Anatomy and Physiology, The University of Melbourne, Parkville, Victoria 3010, Australia
[8] Department of Biomedical Engineering, The University of Melbourne, Melbourne, Victoria 3010, Australia
[9] School of Physics, The University of Melbourne, Parkville, Victoria 3010, Australia

Corresponding Author: Wei Tong −Email: wei.tong@unimelb.edu.au

ABSTRACT:
This study demonstrates the control of neuronal survival and development using nitrogen-doped ultrananocrystal- line diamond (N-UNCD). We highlight the role of N-UNCD in regulating neuronal activity via near-infrared illumination, demon- strating the generation of stable photocurrents that enhance neuronal survival and neurite outgrowth and foster a more active, synchronized neuronal network. Whole transcriptome RNA sequencing reveals that diamond substrates improve cellular−substrate interaction by upregulating extracellular matrix and gap junction-related genes. Our findings underscore the potential of conductive diamond as a robust and biocompatible platform for noninvasive and effective neural tissue engineering.

# 1. INTRODUCTION

Techniques that can enhance neurite development and outgrowth hold great potential in tackling both neural injuries and neurodegenerative diseases by fostering neural regener- ation and restoration.[1] Recent years have seen the develop- ment of a variety of approaches aiming to modulate neuronal survival and neurite outgrowth via chemical, electrical, and mechanical stimuli.[2] As neurons are electroactive, electrical stimulation has been considered to be one of the most efficient methods. The therapeutic potential of electrical stimulation has already been demonstrated in the treatment of peripheral nerve injuries, where it promotes and guides neurite outgrowth.[3,4]

A range of photoelectric materials have been developed and employed to modulate neural activities by converting light into electrical signals.[2,5] Compared to direct electrical stimulation, optical stimulation via these photoelectric materials provides several advantages.[6] First, optical stimulation eliminates the need for wires that connect electronic implants with external devices for powering and data transmission. The potential for wireless stimulation simplifies the device design and fabrication, reduces surgical complexity, and decreases the risks of post-surgery infection by eliminating percutaneous connectors. Second, in electronic implants, the spatial pattern of electrical stimulation is predefined by the electrode array arrangement, and the electrode density limits its resolution. However, optical stimulation offers higher flexibility and precision; advanced optical instruments allow for the delivery of versatile illumination patterns at a high spatiotemporal resolution. Third, many photoelectric materials respond specifically to different illumination wavelengths, which leads to diverse cell/material interactions. Therefore, these photo- electric materials can provide an additional degree of freedom for control of neuronal activities, depending on the chosen wavelength and spatial arrangement of the illumination.

Some of the photoelectric materials that have been proposed and employed for controlling neuronal development and neurite outgrowth include silicon,[7] conductive polymers such as poly(3-hexylthiophene-2,5-diyl) (P3HT),[8–16] and carbon-based materials such as graphene oxide[17–19] and graphitic carbon nitride.[20–23] The favorable neural stimulation effects of these materials under illumination have been demonstrated using various cultured cell models, including pheochromocytoma (PC12) cells,[9,10,20,21] primary cortical neurons,[12,14] and neural stem cells.[8,17–19] Most of these photosensitive materials generate Faradaic currents upon illumination via photoinduced electrons and holes.[20,21] Previous research indicates that the direct transfer of electrons (Faradaic charge injection) may have a negative impact on the survival of neurons due to the production of a high level of reactive oxygen species (ROS).[9] Therefore, additional chemicals, such as ascorbic acid, are required to scavenge the photoexcited holes in the cultures,[20,21] which limits their *in vivo* and clinical applications. Therefore, photoelectric materials that do not produce a Faradaic charge injection are highly desirable. In general, biomedical applications can benefit from using light with longer wavelengths due to their deeper tissue penetration potential and reduced phototoxicity, but most photoelectric materials respond only to visible light. Finally, the long-term stability of the implant materials is an important parameter to consider when designing biomedical devices. However, the long-term stability of photoelectric materials has rarely been studied and reported.[16]

Diamond has been considered a suitable material for many biomedical applications due to its high chemical/biochemical inertness and durability.[24,25] These biomedical applications have been reviewed,[24–26] with *in vitro* applications including cell culture substrates and *in vivo* applications focusing more on implant and prosthesis coatings. The biocompatibility of diamond has been confirmed in many studies.[24,25,27] For example, the number of human sperm cells with grade A motility was found to increase by ∼300% after 1 h contact with the nanodiamond coated surface, as compared to polystyrene Petri dishes.[27] Nanostructured diamond has also been shown to facilitate the growth of cortical neurons[28] and to promote the attachment, proliferation, and survival of neural stem cells.[29] Among different types of diamond, nitrogen-doped ultrananocrystalline diamond (N-UNCD) stands out due to its high electrical conductivity, which is a result of nitrogen additives increasing the concentration and thickness of grain boundaries that are mainly composed of $sp^2$-bonded carbon species.[30,31] Moreover, the surface of N-UNCD can be engineered to exhibit various electrochemical properties.[32–35] The biocompatibility of N-UNCD has been validated both *in vitro*[32] and *in vivo*.[36,37] N-UNCD has already been employed successfully as an electrode material for neural stimulation and recording,[38–41] including its application in a high-acuity retinal implant for vision restoration.[36,42]

In our previous work, we explored the photoelectric properties of N-UNCD, highlighting its response not only to visible light wavelengths but also to the near-infrared (NIR) range.[33,34,43] We observed that under NIR illumination, the photoresponse changes depending on the surface termina- tion.[34] For example, in saline solution, hydrogen-terminated N- UNCD shows an illumination induced Faradaic charge transfer mechanism, while oxygen-terminated N-UNCD electrodes predominantly exhibit capacitive photocurrents. This capaci- tive charge transfer involves a redistribution of charged chemical species within the electrolyte without any direct charge exchange across the electrode/electrolyte interface.[44,45] Such a capacitive charge transfer is considered safer than the Faradaic charge transfer for biological applications.[46] In a recent study, we demonstrated oxygen annealing as an effective method to enhance the capacitive photoelectric responses of N-UNCD to NIR illumination.[33] Our findings indicated that photoresponses increased proportionally to the annealing duration, reaching a maximum response when heated in oxygen at 420 °C for 25 h.[33]

In this study, we investigate the use of oxygen-annealed N- UNCD as a robust and effective photoelectric platform for wirelessly modulating neuronal development and neurite outgrowth. To achieve this, we first fabricated oxygen-annealed N-UNCD electrodes with different degrees of photorespon- sivity to NIR. We then characterized their surface properties using photoelectrochemistry, X-ray photoelectron spectrosco- py (XPS), scanning electron microscopy (SEM), and atomic force microscopy (AFM). Following this, we assessed the long- term stability of their photoresponses under different biological conditions. We then assessed the impact of these materials on neuronal development and neurite outgrowth using primary cortical neuron cultures, as depicted in Scheme 1. Finally, we investigated the effects of substrates and stimuli on cells via calcium imaging and whole transcriptome RNA sequencing. This approach allowed us to evaluate not just the structural but also the functional and molecular effects of the oxygen- annealed N-UNCD platform on neuronal development and neurite outgrowth.

## 2. RESULTS

**2.1 Diamond Fabrication and Surface Characterization.** N-UNCD films were deposited on silicon substrates using microwave plasma-enhanced chemical vapor deposition, as detailed in the Experimental Section. Oxygen annealing time affects the electrochemical and surface properties of the N- UNCD samples. We investigated N-UNCD samples after 3 annealing durations: 5, 15, and 25 h (hereafter denoted 5-h OA, 15-h OA, and 25-h OA, where OA refers to "oxygen anneal"). Figure 1a shows the photocurrents of the N-UNCD samples measured in a saline solution while pulsing an 808-nm laser at an intensity of 41 mW mm$^{-2}$ and a frequency of 50 mHz. Light with an 808-nm wavelength has been previously demonstrated capable of penetrating human scalp, skull, meninges and brain, reaching a depth of approximately 40 mm.[47] Compared to the 660- or 940-nm wavelengths, 808 nm showed less absorption and scattering.[47] Consistent with our previous findings,[33] the 5-h OA sample showed a typical capacitive charge transfer in response to pulsed illumination. This is evidenced by a sharp initial peak in the photocurrent followed by a slower decay with a charge-balanced and biphasic waveform. For samples treated with 15- and 25-h oxygen annealing, we observed photocurrent transients that exhibit both capacitive and Faradaic characteristics. The presence of positive and negative currents in response to the light being turned on and off indicates a predominant capacitive behavior. However, the asymmetry of these currents combined with the residual current at the end of the positive pulses suggests the presence of a minor Faradaic current component. This is in line with previous modeling work, which found that both capacitive and Faradaic currents are present for longer annealing times.[16] The amplitude of the peak photocurrent demonstrated an increase in correlation with the annealing duration, ultimately reaching a maximum of 96.15±3.45 $\mu A/cm^2$ in the samples of N-UNCD that underwent a 25-h annealing process.

We previously suggested that oxygen annealing improved the photoresponse of N-UNCD via a combination of the introduction of oxygen surface functionality and grain boundary etching.[33] In our previous research, we utilized Raman spectroscopy, near-edge x-ray absorption fine structure (NEXAFS) measurements, and electrochemical equivalent circuit fitting to elucidate the varying photoresponses of diamond samples.[33] Figure 1b displays the high-resolution carbon 1s XPS spectra of the N-UNCD samples after different annealing periods. An increase in binding energy indicates that more oxygen functional groups were chemically bonded to the surface when the annealing duration increased.19 The surface morphology of the N-UNCD samples was expected to change due to grain boundary etching. The SEM images of the N- UNCD samples are shown in Figure 1c, which did not show a significant change in surface morphology. However, AFM (Figure 1d) revealed that the root-mean-square (RMS) roughness of the surface decreased somewhat with annealing duration. The oxygen annealing additionally increased the hydrophilicity of the samples, resulting in smaller water contact angles on the samples that underwent a longer oxygen annealing duration (Figure 1e).

**2.2 Diamond Photoelectrochemical Stability.** Electrode stability is important for long-term applications; therefore, we investigated the stability of the N-UNCD photoresponse under different conditions. We first compared their photoresponse before and after sterilization, an essential procedure before using the samples in cell studies (Figure 2a,c). Sterilization was conducted using an autoclave and at 134 °C and 235 kPa for 7 min. All of the samples showed high stability, with only a less than 6% increase in photocurrents after the autoclave treatment. We then incubated the autoclaved samples for 3 weeks in a culture medium containing 15% horse serum at 37 °C and 5% $CO_2$. This horse serum-supplemented culture medium has been used previously to mimic physiological conditions.[48] It is therefore used here to investigate the possible effect of proteins in solution on the surface termination and photoresponse. The photocurrent remained stable on the 5-h OA and 15-h OA N-UNCD samples but dropped to 10.7% for the 25-h OA N-UNCD samples. The shape of the transient photocurrent curve did not change and stayed stable (Figure 2c). The stability of the electrodes was further demonstrated by measuring the photoresponse after repetitive laser pulses. Figure 2b summarizes the change of photocurrents after the samples were exposed to over 3000 pulses at a frequency of 50 mHz, for a total of 16.5 h. The illumination was performed at an intensity of 41 mW mm$^{-2}$. The measurement was repeated every 150 pulses, and the peak photocurrents showed less than 7% fluctuation in all the samples but remained relatively stable across the testing regime. The transient traces from one 15-h N-UNCD sample after different pulses are shown in Figure 2d, illustrating the stability of this sample after a long period of pulsing.

**2.3 Effect of Diamond and Light Exposure on Neuronal Structures.** To evaluate the impact of photo- response from the N-UNCD electrodes on neuronal cells, we built a light exposure device, as depicted in Figure 3a. Primary cortical neurons were harvested from postnatal rats and seeded onto three types of N-UNCD samples placed in 24-well dishes with standard Neurobasal-A medium supplemented with B-27, Glutamax, and Penicillin-Streptomycin. We used standard coverslips as the control for comparison. Primary cultures, preferred over immortal neuronal cell lines, are more likely to recapitulate the properties of neuronal cells *in vivo*.[49] Prior to cell seeding, all samples were preincubated in the Neurobasal- A medium, as described in the Experimental Section, with no additional layer to promote adhesion. Illumination occurred the day after seeding to avoid confounding effects on cell attachment. Half of the cultures were subjected to overnight exposure to 810-nm LEDs at an intensity of 20 mW cm$^{-2}$ and a frequency of 50 mHz, for a duration of 16 h. The chosen LED intensity was deemed to be the maximum level that would not introduce a significant temperature change (within 0.5 °C).

Following this exposure, the cells were then fixed and immunostained on Day 4 for morphological analysis. The representative images of neurons on different substrates, both with and without NIR exposure, are shown in Figure 3b. On all sample types, neurons survived, extended neurites, and formed dense networks. For each condition, three replicates were conducted with imaging data collected from a minimum of five random regions from each individual sample. Statistical analysis was performed to further study the survival rate and neurite development by comparing cell density (determined by propidium iodide staining), neuron coverage (quantified as the area percentage represented by β-III tubulin staining), and the number of neurites per neuron (manually enumerated using ImageJ), as illustrated in Figure 3c−f. We also studied the cell condition on each substrate by determining the proportion of neurons that either remained isolated (cell bodies not in contact) or formed small clusters (two neurons with adjacent cell bodies), as shown in Figure 3f. This assessment was conducted to investigate the impact of the substrate surface on cell aggregation. Neurons prefer to adhere and spread onto surfaces that offer optimal biochemical and biomechanical cues.[50] Therefore, a higher percentage of single- or dual-cell aggregations suggests an advantageous condition, as it indicates the cells' preference to adhere to the substrate over clustering with other cells.

On Day 4, neurons on all the N-UNCD electrodes exhibited better survival rates compared to those on the coverslip controls. This was evident through the significantly higher cell densities, neuronal coverage, and the increased count of neurites per neuron (Figure 3c−f). In the absence of light exposure, the survival rates of cells on the N-UNCD samples demonstrated an increase in correlation with the annealing duration (Figure 3c). This was most noticeable in the 25-h OA samples, which presented the highest density and coverage of neurons. This result aligns with previous findings[32] and could be attributed to the combined impact of the additional surface oxygen functional groups and changes in surface roughness and hydrophilicity. These aspects are discussed in more detail in Section 3.1. The cell aggregation did not appear to be sensitive to the degree of annealing of the N-UNCD samples (Figure 3f).

Neurons adhered to the N-UNCD samples, and the controls showed remarkably different responses to NIR stimulation (Figure 3c−f). For neurons adhered to the control coverslips, light exposure had no significant effect on the cell density, neuron coverage, or the number of neurites per neuron, further validating the safety of the light exposure conditions applied in this study. Likewise, the cell survival rate and neurite development on the 5-h OA samples did not show significant differences upon exposure to light. In contrast, light exposure significantly enhanced cell survival and neurite development on the 15-h OA surfaces. Following light exposure, the cell density, the neuron coverage, and the number of neurites per neuron on the 15-h OA electrodes increased by 16%, 18%, and 7%, respectively (Figure 3c−e). Moreover, a larger percentage of single-cell or dual-cell aggregates was observed (Figure 3f), further affirming the advantageous effect of light exposure on the 15-h OA samples. Interestingly, light exposure inhibited neuronal survival and development on the neurons adhered to the 25-h OA samples, leading to a significantly reduced cell density, lower coverage, and fewer neurites. However, the cell clustering condition remained unchanged. The cells on the 15- h OA samples exposed to light displayed a similar density, coverage, and neurite number to the 25-h OA samples without light exposure. Still, the aggregation analysis (Figure 3f) revealed that cells exhibited a preference for the former condition, tending to refrain from forming large aggregates.

Upon validating the beneficial impact of the photoresponse from the 15-h OA samples using short-term cultures (4 days), we extended our investigation to analyze the properties of the neurons after a longer culture period (2 weeks). We examined the morphology of the neurons via immunostaining to evaluate the survival rate of the cultures on Day 14 (Figure 3g,h). Consistent with the short-term culture findings, cells survived better on the 15-h OA samples compared to the coverslip controls, demonstrated by a significantly higher cell density and neuron coverage. NIR exposure further enhanced the survival rate and neurite development on the 15-h OA samples but not on the coverslip controls. The cell density on the 15-h OA samples increased by 1.4 times, and the neuron coverage increased by 27% as a result of light exposure.

**2.4 Probing Calcium Activities in Response to Diamond and Light Exposure.** We investigated the intracellular calcium activities and their neuronal network properties under different conditions via calcium imaging on Day 14. Figure 4a shows an image of neurons loaded with calcium indicator Fluo-4 AM on a 15-h OA N-UNCD sample following light exposure on Day 14. These labelled neurons exhibited spontaneous activities with corresponding calcium transients shown in Figure 4b. The fluorescence changes, $\Delta F/F_0$, of individual ROIs were analyzed by using CALIMA software, allowing for the identification of calcium spikes. The representative images and recordings of neurons loaded with Fluo-4 AM on different substrates are shown in Figure S1 and Video S1, respectively. Each spike displayed an onset due to neuronal activation, followed by a decay back to the baseline, which corresponded to the slow unbinding rate of calcium ions from the fluorescent probe. Therefore, each peak in the traces represented a single peak or a burst of intracellular calcium activity.

Upon identification of neuronal activities, the analysis proceeded to examine the number of active neurons (Figure 4c). Given the larger number of surviving neurons on the 15-h OA samples, the number of active neurons was significantly higher on these samples compared to the coverslip controls (Figure 4c). NIR exposure led to a 27% increase in active cell density on 15-h OA samples. Further analysis of the ratio of active cells to the total number of cells also revealed higher values on the 15-h OA samples than the coverslip controls (Figure 4d), but light exposure did not have any significant impact.

Pearson's cross-correlation analysis, facilitated by CALIMA software, was utilized to determine the degree of spatial correlation among neurons exhibiting spontaneous calcium activities. Pearson's algorithm considers the timing of the spikes across ROIs and reports a score between +1 and −1, with +1 symbolizing perfect correlation between the distribution signals, a score of −1 representing inverse distribution, and 0 signifying uniform distribution. Pearson's scores, summarized in Figure 4e, were notably higher on the 15-h OA samples compared to the coverslip controls. On the N-UNCD samples, light exposure improved the average Pearson's score from 0.7 to nearly 0.8, implying enhanced rhythmic activities in the neurons and a more mature neuronal network. This observation aligns with the morphology findings in Figure 3g,h, which revealed higher connectivity.

**2.5 Transcriptional Profile Changes Induced by Diamond and Light Exposure.** We examined the tran-scriptomic profiles of cells cultured on both control and 15-h OA samples, with or without light exposure, on Day 14. RNA sequencing was performed using cells from each condition across three replicates (Figure 5). Using integrated Differential Expression and Pathway (iDEP) analysis,[51] we first performed a principle component analysis to investigate similarities between gene expression profiles across various samples, as represented in Figure 5a,b. Notably, the transcriptional profile of cells cultured on the 15-h OA N-UNCD samples clearly diverged from the control samples and formed a distinct cluster in Figure 5b. Principal Component 3 (PC3) exhibited a strong correlation with the substrate type, with a variance value of 13% and a statistically significant $p$-value of $1.39 \times 10^{-6}$. Thus, the N-UNCD substrates lead to changes in the transcriptional profiles of neurons that may drive their improved survival and activity.

Next, a differential expression analysis was carried out to identify the quantity of differentially expressed genes across various conditions (Figure 5c−f). The cutoff threshold for analysis was set at $p = 0.05$ and Fold Change = 1.5. Figure 5c summarizes the numbers of differentially expressed genes in each pair, and a comprehensive list of the differentially expressed genes can be found in Table S1. We identified more differentially expressed genes when comparing 15-h OA to control samples without light exposure than between light-on and light-off conditions, regardless of substrate type. This suggests that the substrate type had a more significant impact than light exposure on the transcriptional responses of the cells.

In comparison to control samples without light exposure, the diamond substrate significantly upregulated several genes related to critical neuronal functions (Figure 5d). These genes include Gjb6 and Gjb2, which are associated with gap junction proteins.[52,53] Gap junctions provide critical pathways for the propagation of presynaptic electrical currents to postsynaptic sites in neurons, thereby playing a vital role in signal transmission.[54] Additionally, diamond substrates in- duced the differential expression of genes involved in sodium channel activities, including Slc22a6, Slc6a13, Scn7a, Scl30a3, Slc22a3, and Slc25a34. The collective activity of these genes could potentially facilitate a more active neuronal network on the diamond, as supported by our calcium imaging experi- ments (Figure 4). Diamond substrates were also observed to upregulate genes involved in cell attachment, including Cdh5, a gene encoding calcium-dependent cell adhesion molecule,[55] and Col13a1, associated with collagen production.[56] Collagen, an extracellular matrix component, is known to regulate neuronal attachment and signal transduction.[57] Focusing on the differentially regulated genes, we performed a Gene Ontology (GO) enrichment analysis using iDEP.[51] The GO terms that reached statistical significance and ranked in the top 10 are listed in Figure 5g and Table S2 and are categorized into three classifications: biological process, cellular compo- nent, and molecular function. Since gap junction genes, Gjb2 and Gjb6, were significantly upregulated on diamond compared to control samples, they resulted in the enhance- ment of gap-junction-mediated intercellular transport in the biological process category (Table S2). Similarly, "gap junction channel activity involved in cell communication by electrical coupling" was enriched within the molecular function category. These genes led to the upregulation of the connexin complex, a critical element in forming gap junctions. Extracellular space was the most significantly upregulated GO cellular component term, suggesting that diamond may enhance interactions with the cells via the extracellular matrix. Furthermore, genes such as Col13a1 and Cdh5 also induced the upregulation of several substrate-binding processes. These processes potentially contribute to improved cell adhesion and survival on diamond substrates.

When studying the effects of light exposure, Fosb emerged as the only differentially expressed gene on both the control and 15-h OA samples due to light exposure (Figure 5e,f). Significant GO terms ($p = 0.01$) resulting from light exposure are summarized in Figure S2 and Tables S3 and S4. For the control samples, no GO signals pertinent to neuronal health and behavior emerged. Comparing light-on and light-off conditions on the diamond samples at Day 14, only 10 genes were differentially expressed, suggesting that GO enrichment analysis may not provide reliable insights. Never- theless, the higher number of significant GO terms observed in diamond samples might indicate a more specific and directed effect of light exposure on cells when they are coupled with diamond substrates.

## 3. DISCUSSION

**3.1 Substrate Impact.** In this work, we engineered N- UNCD electrodes demonstrating different levels of photo- electric effects and investigated the response of neurons attached to these surfaces upon exposure to NIR light. Consistent with our previous findings, all N-UNCD samples displayed a predominantly capacitive photoresponse to NIR stimulation. An increase in oxygen annealing time led to enhanced photoresponsivity in N-UNCD electrodes, which is likely due to an increase in surface oxygen functional groups (Figure 1b) and the etch of grain boundaries (Figure 1d). Our photoelectrochemical stability (Figure 2) showed that all of the samples remained highly stable under a variety of biologically relevant conditions. This indicates their potential suitability for robust, long-term biomedical applications.

Our study showed that neurons survived better on N-UNCD electrodes, regardless of the annealing duration, compared to the coverslip controls, as shown in Figure 3. These results further confirmed the biocompatibility of N- UNCD. Previous studies have consistently shown diamond to be superior to other routinely used substrates for *in vitro* cell cultures.[24,25] For example, macrophages cultured on diamond were found to down-regulate inflammatory cytokines com- pared to tissue culture polystyrene (TCP),[58] suggesting that diamond implants may induce a less severe immune response. Various cells such as epithelial, fibroblasts, and neural stem cells have also been found to attach and proliferate better on diamond than quartz, glass, and TCP.[59−61] Furthermore, Sommer et al. reported a significant increase in the vitality of human sperm cells when in contact with diamond, in contrast to TCP.[27]

Among various diamond samples, the 25-h OA sample was the most effective in enhancing cell survival, with the highest neurite coverage and number of neurites per neuron. Surface properties, such as roughness and wettability, have been shown to play important roles in neuronal survival and growth. We previously suggested surface roughness as a major factor influencing neuronal growth on diamond.[32] The role of topography in neurite outgrowth has been demonstrated on different materials, as neurite initiation requires a certain amount of physical space for the proper orientation of cytoskeletal filaments to sprout new processes.[62] The 25-h OA N-UNCD samples, characterized by the lowest surface roughness (Figure 1d), may enhance neuronal adhesion and neurite outgrowth by facilitating cellular interactions. Fur- thermore, the additional oxygen functional groups on the 25-h OA N-UNCD surface created a more hydrophilic surface, as confirmed by the water contact angle measurements (Figure 1e). Hydrophilic surfaces tend to promote better cell adhesion, which is critical for neuronal survival and growth. These types of surfaces allow for stronger interactions between the surface and the extracellular matrix proteins, which in turn facilitate cell attachment.[63] The extracellular environment *in vivo* is generally hydrophilic due to the presence of water and polar molecules. Therefore, culturing neurons on hydrophilic surfaces more closely mimics their natural environment, potentially leading to better cell behavior and functionality.

In this work, we conducted RNA sequencing to compare the transcriptome profiles of cells adhered to coverslip controls and the 15-h OA samples. To the best of our knowledge, this is the first study examining the transcriptome profile of neurons adhered to diamond substrates. Our results highlight differ- ences in gene expressions between cells on these two types of materials, i.e., diamond and coverslip, which in turn provide insights into the potential underlying mechanisms. Diamond substrates seemingly enhance cell survival by upregulating genes associated with extracellular matrix production, such as Col13a1 and Cdh5, thereby improving cellular adhesion and viability.

In addition to a higher survival rate, our calcium imaging results suggest that diamond substrates foster a larger proportion of spontaneously active neurons and a higher level of synchronized neuronal activities (Figure 4). This observation could be attributed to the upregulation of gap junction protein genes, Gjb6 and Gjb2. Gap junctions play critical roles in signal transmission between neurons by controlling the plastic properties of electrical synapses.[64] However, the influence of biomaterials on gap junction formation remains an understudied area. Our findings suggest that further exploration of gap junctions may shed light on the interplay between biomaterials and cells.

**3.2 Light Exposure Impact.** Low-intensity red or near- infrared light, has been used as a therapeutic approach to stimulate, heal, regenerate and protect injured or degenerated tissues.[65] This therapeutic approach, also known as photo- biomodulation, was initially primarily studied for stimulation of wound healing, but its beneficial impacts have been found useful in treating brain disorders as well. However, the precise cellular mechanisms underpinning photobiomodulation are not yet fully understood. Known effects of red or near-infrared light include its absorption by cytochrome C oxide and other molecules, which stimulates ATP synthesis and affects ROS production.[65] In our study, we did not notice any significant impact of 810-nm illumination on the structure and calcium activities of neurons adhered to coverslip controls. The RNA sequencing study revealed some differentially expressed genes in cells on coverslip controls with or without light exposure, but there was no enrichment for pathways clearly related to neuronal health or survival. This indicates that the impact of light exposure on cells on control samples is not highly specific, implying that the observed effect may not be attributed to a well-defined biological process or pathway. However, it is well- established that cells may react differently to light, depending on the dose and frequency.

The effects of light stimulation have been published on cells adhered to various photoelectric materials.[5–14,17–23] Most previous studies used a single material, only examined cellular responses to one level of stimulation, and only reported the favorable effects of light exposure on cells.[9,10,14,17–21] By applying different surface treatments to N-UNCD, we were able to examine the cellular response to varying magnitudes of the photocurrent and their mechanisms. We generally observed that the 15-h OA N-UNCD sample produced the most beneficial effect on cells. This could be attributed to a combination of factors. First of all, the 15-h OA sample exhibited a greater capacitive photoresponse than the 5-h OA sample, which is expected to produce a greater membrane depolarization in the cells. Therefore, the illumination only showed beneficial effects on cells adhered to samples with a medium level of photocurrents (15-h OA N-UNCD) and had negligible influence when the photocurrent was small (5-h OA N-UNCD). On the other hand, although the 25-h OA N-UNCD sample displayed an even higher photoresponse, it also produced a greater proportion of Faradaic photocurrent. This Faradaic photocurrent is known to cause adverse effects in cells[66] and the overdose of photocurrents from 25-h OA N-UNCD negatively impacted neuronal survival and neurite outgrowth. Since illumination did not influence neuronal growth on the coverslip controls, the observed effects on illuminated N-UNCD substrates were mostly attributed to the excitation of the photoactive material. This finding aligns with previous research, which demonstrated varying cellular responses to electrical stimulation under different charge injection magnitudes.[2] For example, Liu et al. showed that a charge injection of 0.08 $\mu$C, in a range of 0.01−0.4 $\mu$C, elicited the strongest osteogenetic response in osteoblast precursor MC3T3-E1 cells.[67] Our results provide evidence that different levels of capacitive and Faradaic photocurrents from N-UNCD electrodes can be used to regulate neuronal survival and development.

In this study, we selected an illumination intensity that minimally raised the temperature of the culture medium by less than 0.5 °C to mitigate potential photothermal effects. However, localized photothermal effects on the substrate surface might occur, which would not be reflected in the overall medium temperature. Further investigations are necessary to discern and quantify the specific contributions of the photothermal and photoelectric effects of the substrates on cell cultures.

Here, we studied the intracellular $Ca^{2+}$ activities of neurons on the 15-h OA N-UNCD substrates and coverslip controls after 2 weeks of culture. The application of light stimulation induced more spontaneously $Ca^{2+}$ active cells on 15-h OA N-UNCD samples. Previous studies have demonstrated the crucial role of intracellular $Ca^{2+}$ activities in cellular response to photoelectrical stimulation.[8,10,14,15,22] For example, Wu et al. reported that photostimulation using P3HT nanoparticles could elevate $Ca^{2+}$ levels and increase the expression of L-type voltage-gated calcium channels (L-VGCC) in bone marrow mesenchymal stem cells (BMSCs).[14] They further confirmed the significance of $Ca^{2+}$ by showing that the calcium inhibitor $GdCl_3$ could inhibit the effects of photostimulation on neuronal differentiation of BMSCs.

Contrary to expectations based on previous studies, our RNA sequencing analysis did not reveal significant changes in the expressions of L-VGCC related genes in cells on the 15-h OA N-UNCD samples after light exposure. This suggests that the different behavior observed in neurons on the diamond substrate, with and without light exposure, might be attributed to a different mechanism. This hypothesis is further supported by prior research that investigated the combined effect of light illumination and diamond on sperm cells. The study found that 670-nm LED illumination enhanced the motility of sperm cells on nanodiamond surfaces.[27] with light exposure suggested to increase ATP levels in cells while the nanodiamond surface functioned as a scavenger for the harmful ROS triggered by light exposure. Similarly, our findings provide a hint, though tentative, that light exposure may downregulate cellular responses to oxygen-containing compounds (Figure S2). This could, in theory, make cells less susceptible to ROS stimulation, but additional research is needed to confirm this possibility.

**3.3 Limitations.** This study has several limitations that can be addressed in future research. First, our study has yet to comprehensively investigate stimulation parameters such as illumination time, frequency, and wavelength. Our initial studies illuminated only cells overnight. Future work will expand the parameter space, seeking more precise control over neuronal survival and development. Second, while we have provided some insights into the mechanism of how diamond and photoelectrical stimulation modulates neuronal behavior, the exact mechanism remains unclear. Other molecular tools and sequencing strategies could uncover more genes and pathways involved. In this study, light exposure was conducted on Day 2, and our RNA sequencing experiments were performed on cells at Day 14. This means that our experimental design may not have captured transient cellular responses to both light exposure and the substrate. This is supported by our observation that the benefits of both diamond and photoelectrical stimulation were present on Day 14. Future experiments could involve the monitoring of the transcriptome profile of the cells across different time points to better understand how these variables interact. In addition, it is also possible that the beneficial molecular impact of light is at least mediated partially through nontranscriptional mechanisms, as identified previously.[14] These mechanisms could potentially be elucidated through techniques such as reverse transcription polymerase chain reaction (RT-PCR) to analyze gene expression and Western blotting to assess specific protein expression. Third, the present work was performed using cultured cells *in vitro*. Therefore, the practical applications of this platform include the use of N-UNCD as a substrate for *in vitro* neuronal cultures coupled with NIR illumination to expedite culture maturation. Translating these findings to *in vivo* and eventual clinical use will require additional research. N-UNCD films can be applied free- standing by etching away the silicon substrates or used as coatings on other materials or devices.[32] The safety of N- UNCD for chronic applications has been previously demon- strated by implanting N-UNCD adjacent to the retina tissue.[36] As such, a potential application of our findings could involve the use of N-UNCD electrodes in retinal implants, alongside NIR stimulation, to slow down or rescue the degeneration of neurons in retinal diseases such as retinitis pigmentosa and age-related macular degeneration.

## 4. CONCLUSIONS

This work reports, for the first time, the use of photoelectrically conductive diamond to modulate neuronal survival and outgrowth, thereby providing a potential platform for neural regeneration. N-UNCD films, after oxygen annealing, can generate varying levels of photocurrents in response to NIR stimulation. The photoresponse showed remarkable stability under various biological conditions. Our results showed that N-UNCD surfaces promoted neuronal growth without the need for additional promotive substances, supporting its biocompatibility. Furthermore, the amplitude of the photo- current under NIR stimulation was found to differentially regulate neuronal survival and development. This research enabled the development of a biocompatible, nanostructured, and light-sensitive platform that can modulate neuronal survival and development in a wireless, repeatable, and noninvasive manner, without requiring genetic modification. These results suggest great promise for the future application of N-UNCD in long-term, noninvasive neural regeneration.

## 5. EXPERIMENTAL SECTION

**5.1 Diamond Preparation.** The N-UNCD films used in this work were fabricated on silicon substrates (n+-doped, 1000 $\mu$m thick, single-side polished) using an IPLAS microwave plasma-assisted chemical vapor deposition system, as described in our previous work.[35] Briefly, silicon substrates were first seeded with positively charged 4−6-nm nanodiamonds. Over the course of a three day CVD growth, we maintained a gas mixture consisting of 79% argon, 20% nitrogen, and 1% methane (BOC Australia, purity 99.999%). The microwave power was held at 1000 W, with a stage temperature at 850 °C and a gas pressure at 80 Torr. The resultant thickness of the N-UNCD films was approximately 32 $\mu$m as estimated by FIE Nova NanoLab 200 SEM. All N-UNCD samples in this study underwent oxygen termination through annealing in an oxygen environment (∼0.02 L/min) in a vacuum furnace at 420 °C for 5-, 15-, or 25-h, referred to as 5-h OA, 15-h OA, or 25-h OA, where OA denotes "oxygen anneal".

**5.2 Photoelectrochemical Characterization.** The photoresponses of N-UNCD samples to NIR light were studied using photoelectrochemical measurement, as previously described.[33] These experiments were performed in a saline solution (0.15 M NaCl) by using a three-electrode electrochemical cell connected to a potentiostat (Gamry Interface 1000E). The three electrode setup consisted of a platinum disk counter electrode, a Ag/AgCl reference electrode (eDAQ Pty Ltd.), and the N-UNCD sample serving as the working electrode. For the light source, we used a 0.7 W NIR (808- nm) laser diode (Wuhan Lilly Electronics) pulsed with a 10-s on and 10-s off cycle (50 mHz), using a waveform generator (RIGOL DG4062). During the characterization, the peak illumination intensity was 41 mW mm$^{-2}$ and the illuminated area was 1.7 mm$^2$. The photocurrent density was calculated by dividing the peak photo- current by the illuminated area (1.7 mm$^2$). Each reported measure- ment was the average of ten sequential peak photocurrents, and this was done for three samples at each annealing time.

**5.3 Surface Chemistry Characterization.** X-ray photoelectron spectroscopy (XPS) was used to study the chemical composition of the N-UNCD films. Measurements were carried out on a Thermo- Fisher K-Alpha apparatus using an Al K$\alpha$ radiation source with an $E_{photon}$ of 1486.7 eV.

**5.4 Surface Morphology Characterization.** The surface morphology and roughness of the N-UNCD films were assessed by using scanning electron microscopy (SEM) and atomic force microscopy (AFM). SEM images were acquired using an FIE Nova NanoLab 200 SEM under the high vacuum mode at 30 kV. AFM topographic maps were obtained using an Asylum MFP-3D AFM instrument with a Tap300 Al-G tip in tapping mode. A scanning area of $0.5 \times 0.5$ $\mu$m$^2$ was selected to capture the changes in the surface morphology over nanoscale lengths.

**5.5 Contact Angle Measurement.** The sessile-drop contact angles of the 5-, 15-, and 25-h OA N-UNCD films were measured using a 2.5 $\mu$L droplet of Milli-Q water (Merck Millipore, VIC, Australia). This droplet size was selected to prevent the liquid from touching the edges of the samples. The droplet was carefully placed on the sample surface using a micropipette. Digital images were captured and analyzed using a Drop Shape Analyzer DSA25 (Kruss GmBH, Hamburg, Germany) at room temperature with a Young Laplace fitting method. The contact angle was acquired 100 times for each drop, and the mean contact angle along with the standard deviation was calculated.

**5.6 Stability Tests.** The stability of the N-UNCD photo- responses was evaluated via three different tests. First, samples underwent sterilization using an autoclave before cell seeding, and photoresponses were measured both before and after sterilization. Autoclaving was performed at 134 °C and 235 kPa for 7 min. The second test was designed to assess the stability of the electrodes under physiological conditions. This involved monitoring the photoresponses of the N-UNCD films over time, while they were incubated in a cell culture medium supplemented with 15% horse serum. The medium comprised Neurobasal A with 2% B-27 supplement (Gibco), 2 mM Glutamax (Gibco), and 100 mg/mL penicillin-streptomycin (Gibco). The incubation conditions were maintained at 37 °C, 5% $CO_2$, for a three-week period.[48] Lastly, a third test was conducted to evaluate the stability of the electrodes under conditions of repetitive illumination. The samples were exposed to a pulsed NIR laser operating at 50 mHz with a 10-s on and 10-s off cycle. The peak intensity of the illumination was 41 mW mm$^{-2}$ over 3000 pulses (equivalent to a total duration of 16.5 h). Photoresponses were measured after every 150 pulses.

**5.7 Cell Culture.** All cell culture procedures were performed in compliance with the guidelines of the National Health and Medical Research Council of Australia (Australian Code for the Care and Use of Animals for Scientific Purposes) and approved by the Animal Ethics and Welfare Committee of the University of Melbourne (Ethics ID 1814396). The N-UNCD films and coverslip controls were sterilized via autoclaving and then placed in 24-well dishes for cell culture experiments, with three replicates for each experimental condition. Before cell seeding, all samples were incubated in the Neurobasal A medium (Gibco) for at least 24 h without additional coatings. Primary cortical neurons were harvested by isolating the cerebral cortices from P0−P1 rat pups, as previously described.[32] Briefly, the animals were decapitated, and the heads were immersed in Hank's balanced salt solution (HBSS; Gibco). After the skin and skull were removed, a small section of the cortex was taken off using fine forceps. The meninges were then removed, and the tissue was dissected using scalpel blades and collected. The neurons were obtained from the dissociated tissue by protease digestion for 20 min at 37 °C in HBSS containing 10 $\mu$g/mL DNase I (Sigma) and 250 mg/mL trypsin (Sigma). Soybean Trypsin Inhibitor (Sigma) containing 10 $\mu$g/mL of DNase I was used to stop trypsinization. Neurons were centrifuged and then triturated using a P1000 pipet. A cell culture medium containing Neurobasal A with 2% B-27 supplement (Gibco), 2 mM Glutamax (Gibco), and 100 mg/mL penicillin-streptomycin (Gibco) was used to dilute the triturated cells. The cells were seeded at a density of $1.8 \times 10^5$ cells/well on both diamond films and control samples. Finally, the cultured primary cortical neural cells were incubated at 37 °C in a 5% $CO_2$ environment. The culture medium was changed on the second day (Day 2) and then twice weekly thereafter.

**5.8 Cell Culture Light Exposure.** Half of the samples were exposed to NIR on Day 2 for a duration of 16 h. During the illumination process, the 24-well dish was placed in a custom-built box, and a board fitted with six 810-nm LEDs (OSLUX, SFH 4780S) was situated above the dish (Figure 3a). The LEDs were driven using a GW Instek GPS-1850D power supply and pulsed using a waveform generator (RIGOL DG4062) at 50 mHz with a 10-s on and 10-s off cycle. All samples were subjected to a peak intensity of 20 mW cm$^{-2}$ as measured using a laser power meter (Coherent, Fieldmax). A fan was attached to the LED board to improve heat dissipation. The temperature change monitored during the light exposure was noted to be less than 0.5 °C.

**5.9 Fixation, Immunostaining, and Imaging.** The cellular morphology was assessed on days 4 and 14 through a process of fixation and immunostaining. The samples were fixed in 4% paraformaldehyde solution for 10 min and then in cold (−20 °C) methanol for 10 min, followed by a triple rinse in phosphate-buffered saline (PBS). Subsequently, the samples were incubated for 20 min with a primary antibody solution (mouse anti-$\beta$-III tubulin; Promega) after a 30-min pretreatment with a blocking solution containing 2% fetal calf serum and 2% normal goat serum in PBS. Postincubation, the films were once again rinsed with PBS and then incubated with secondary antibody (Cy3-conjugated goat antimouse immunoglobu- lin; Jackson Immunolabs) and propidium iodide (PI, Sigma-Aldrich) for another 20 min. Finally, the samples were thoroughly rinsed with PBS three times and stored at 4 °C prior to imaging. Random regions of 636 × 636 $\mu m^2$ were selected from each sample and imaged using a confocal microscope (Olympus, FV1200) with excitation lasers at 473 and 543 nm through a Nikon Plan Apo 0.75-numerical aperture (NA) 20× objective. Images were then analyzed using Fiji (ImageJ),[68] MATLAB R2020b and OriginPro 2021.

**5.10 Calcium Imaging.** To study the electrophysiological properties of the cells, calcium imaging was performed on the 15-h OA and control samples on Day 14. On the day of imaging, the cultured cells were washed with PBS and then incubated at 37 °C with a solution containing 5 $\mu$M Fluo-4 AM (Invitrogen) in Ames' medium (Sigma) for 20 min. Following this, the samples were transferred under an upright confocal microscope (Olympus, FV1200) for imaging. To maintain physiological conditions during the imaging process, the cells were continuously perfused with carbonated Ames' solution at a flow rate of 3 to 8 mL/min at room temperature. Images were taken using a Nikon Plan Apo 40× objective (0.8 NA), with an excitation wavelength of 473 nm. At least five different regions from each sample were imaged, each with a field of view of 159 × 159 $\mu m^2$. The images were captured at a frame rate of 4.8 Hz for a total duration of 3 min. Calcium imaging data were processed using the semiautomated open-source $Ca^{2+}$ imaging analyzer CALIMA according to its instructions.[69] This software determined the regions of interest (ROI), and the fluorescence changes in ROIs were normalized to the baseline ($\Delta F/F_0$) and then analyzed. Calcium spikes were detected by setting a threshold and the connections between pairs of neurons were determined via the Pearson cross-correlation analysis. Data were analyzed further using MATLAB R2022b and OriginPro 2021.

**5.11 RNA Sequencing.** On Day 14, cells from 15-h OA and controls were lysed, and their total RNA was extracted in accordance with the manufacturer's instructions using the Qiagen RNeasy Lipid Tissue Mini kit (Qiagen, 74804). PolyA purification, library preparation, and 150bp paired end Illumina whole transcriptome sequencing were performed by the Australian Genome Research Facility to a depth of 50 million reads per sample on the NovaSeq platform. Each sample was assigned a unique barcode. Following sequencing, the acquired sequence reads were aligned against a reference genome library using the Spliced Transcripts Alignment to a Reference (STAR) (version 2.3.5a).[70] The chosen reference was *Rattus norvegicus* from NCBI RefSeq version 6, with the accession number

GCF_000001895.5. Subsequently, the counts of aligned sequences were analyzed using integrated Differential Expression and Pathway (iDEP) analysis.[51] We first performed Principal Component Analysis to compare the similarity between cells under different conditions. Following this, differentially expressed genes (DEGs) were identified using the DESeq2 method with a False Discovery Rate (FDR) cutoff set at 0.05 and a minimum fold change of 1.5. Subsequently, an enrichment pathway analysis in DEGs for each selected comparison was carried out according to the Gene Ontology (GO) categories: biological process, cellular component, and molecular function, using an FDR at 0.01. The results were then exported from iDEP and replotted for clearer visualization and interpretation using MATLAB R2022b.

**5.12 Statistical Analysis.** Statistical analysis was performed using IBM SPSS Statistics 27. One-way analysis of variance (ANOVA) or two-way ANOVAs with posthoc Tukey's tests were performed for multiple comparisons. Data were expressed as mean ± SEM, and $p < 0.05$ was considered statistically significant.

## 6. ACKNOWLEDGMENTS

The authors acknowledge the facilities, as well as the scientific and technical assistance of the Australian Microscopy & Microanalysis Research Facility at the RMIT University. This research was funded by a Medicine/Science Grant from the CASS Foundation (CASS 8612), a Linkage Grant from the Australian Research Council (LP180100638), and a grant from the Australian College of Optometry. A.A. was supported by a Discovery Project from the Australian Research Council (DP210102750). M.B.C. was supported by National Health and Medical Research Council (APP11968410). W.T. was supported by a University of Melbourne Early Career Researcher Grant (2021ECR091) and a Discovery Early Career Researcher Award (DECRA) Fellowship from The Australian Research Council (DE220100302). The authors declare the following competing financial interest(s): S.P. was previously a shareholder in iBIONICS, a company developing a diamond-based retinal implant. S.P., D.J.G., and M.R.I. are shareholders and public officers of Carbon Cybernetics Pty Ltd., a company developing diamond and carbon-based medical device components. A.A. is a shareholder in BrainConnect Pty Ltd., an Australian startup developing physiological and neurophysiological and interven- tional solutions for a range of neurological disorders. The remaining authors declare that the research was conducted in the absence of any commercial or financial relationships that could be construed as a potential conflict of interest.

22

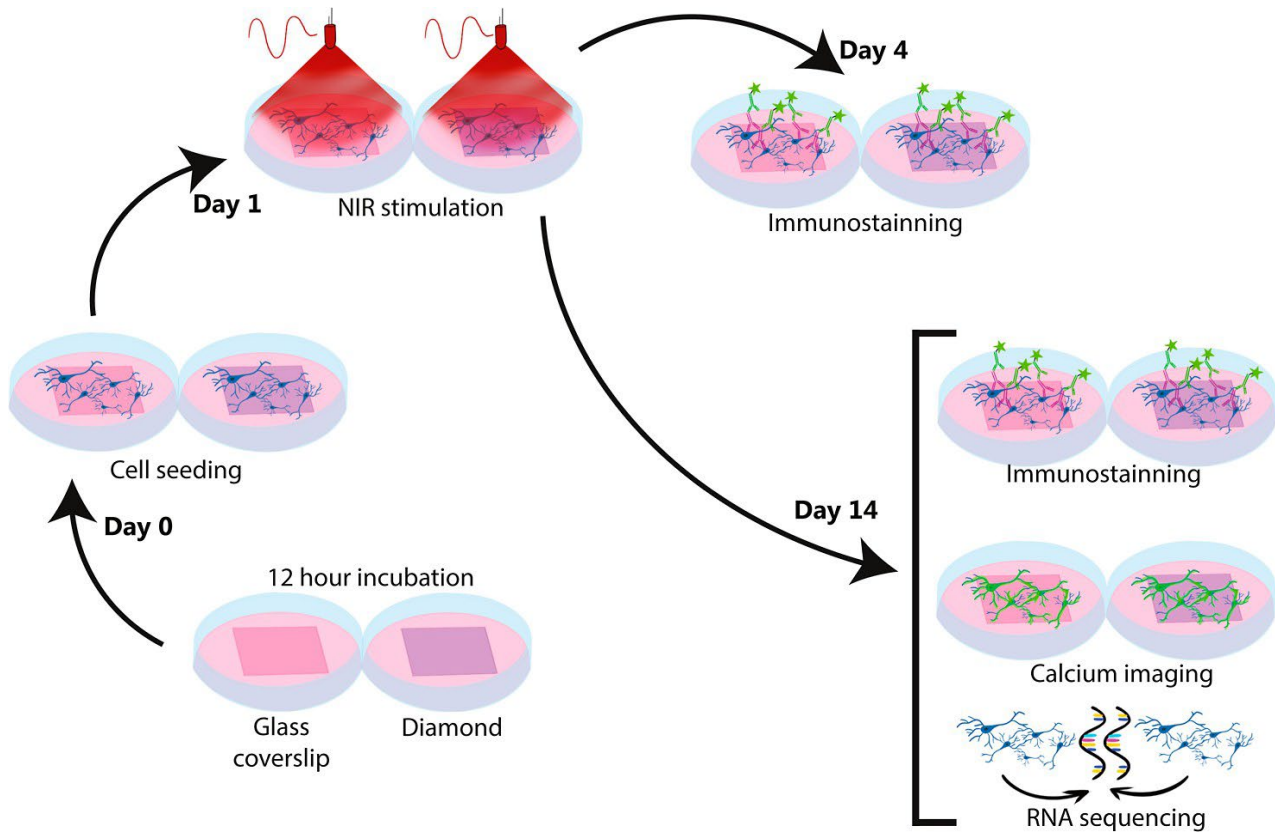

Scheme 1. Methodology Workflow for Studying the Impact of Diamond Substrates and Near-Infrared (NIR) Light Exposure on Primary Cortical Neurons. The process began with incubating glass coverslip controls and diamond substrates in Neurobasal-A medium solution for 12 h. Primary cortical neurons harvested from postnatal rats were seeded onto the samples at a density of $1.4 \times 10^5$ cells cm$^{-2}$. Subsequently, half of the samples from each substrate were exposed to 810-nm NIR light on the first day *in vitro* (Day 1) at an intensity of 20 mW cm$^{-2}$ and a frequency of 50 mHz, for 16 h. Cellular structure and morphology of a portion of the cells were assessed on Day 4 via fixation and immunostaining with *β*-III tubulin and propidium iodide. On Day 14, the assessment was expanded to include not only immunostaining but also calcium imaging and whole transcriptome RNA sequencing. The latter two methods were employed to explore the functional and molecular effects of the substrate and light exposure on the cells.

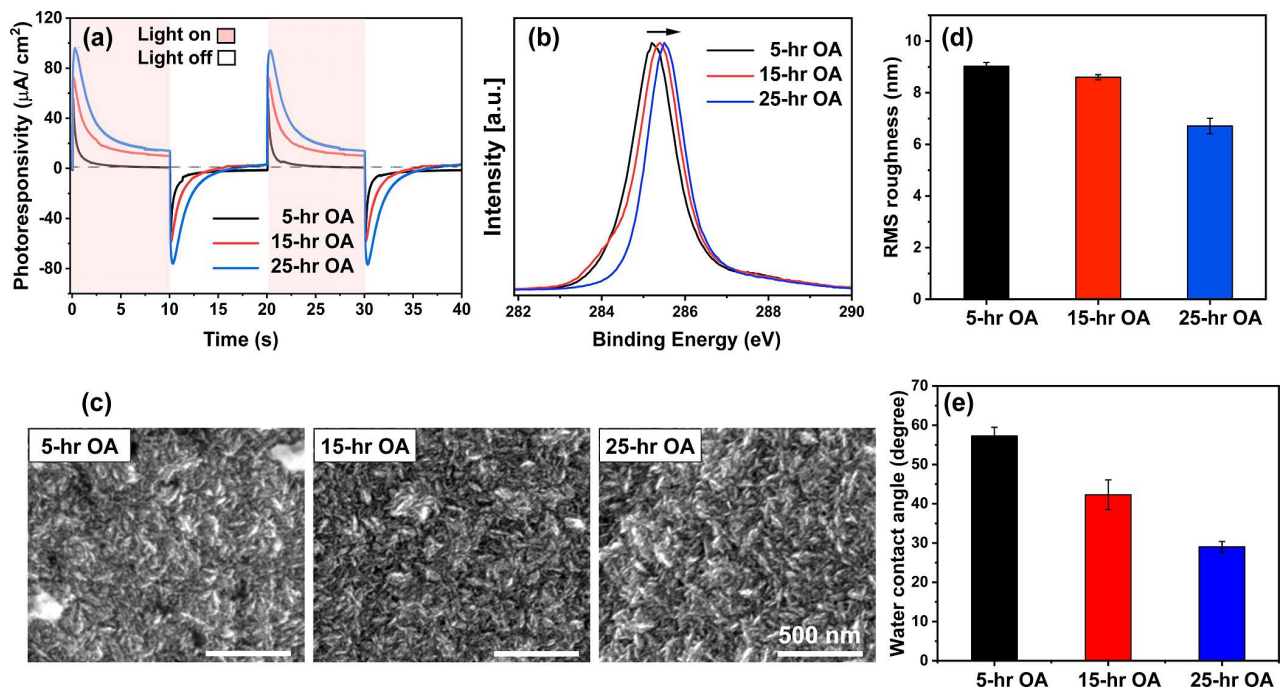

Figure 1. Surface characterization and photocurrent response of oxygen-annealed (OA) N-UNCD samples. (a) Transient photocurrent density of OA N-UNCD samples when subjected to NIR light (808-nm, 50 mHz pulse rate, maximum intensity of 41 mW mm$^{-2}$, and 1.7 mm$^2$ illumination area). The samples were measured in saline solution, using a 3-electrode setup as detailed in the Experimental Section. The peak photocurrent density increases with annealing duration. (b) XPS high-resolution carbon 1s spectra of N-UNCD samples demonstrate an increase in binding energy with annealing duration, suggesting an increase in oxygen functional groups. (c) SEM images show no significant difference in the surface topography of 5-, 15-, and 25-h OA samples. (d) Root-mean-square (RMS) roughness as measured using AFM and (e) water contact angle of N- UNCD samples decrease with annealing duration.

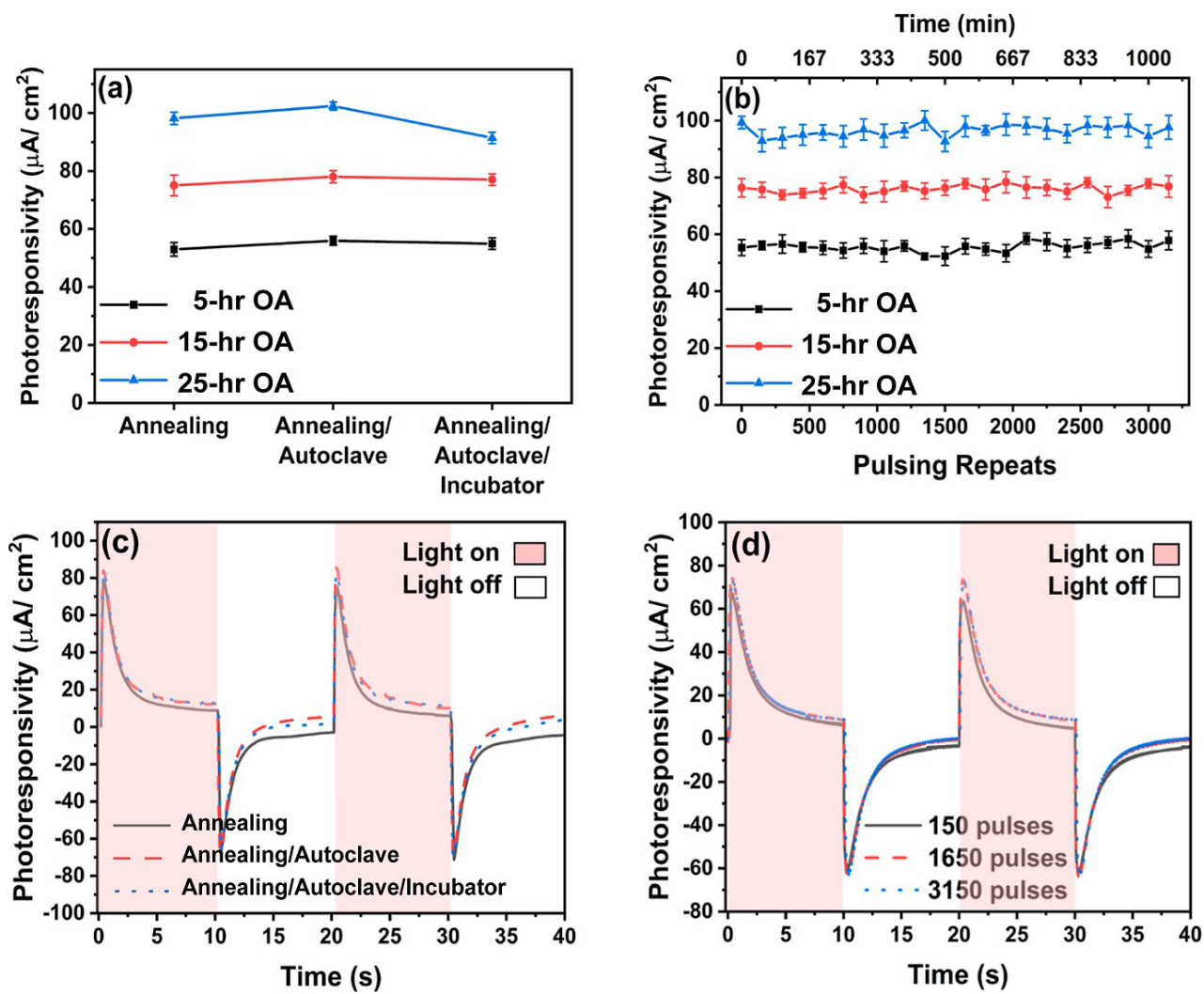

Figure 2. Stability of N-UNCD Photoresponse. (a) Photocurrent comparison for N-UNCD electrode before and after autoclave treatment and 3 weeks of incubation in a cell culture medium supplemented with 15% horse serum. (b) N-UNCD electrode photocurrent fluctuations following repeated, pulsed illumination. The pulsing was repeated 3000 times at a frequency of 50 mHz, and the photocurrents were measured every 150 pulses. Error bars in (a) and (b) represent the standard deviations. (c) showcases the transient photoresponse of a 15-h OA N-UNCD electrode, both preand post autoclave and incubation process. (d) The 15-h OA N-UNCD response following exposure to varied pulsed illumination.

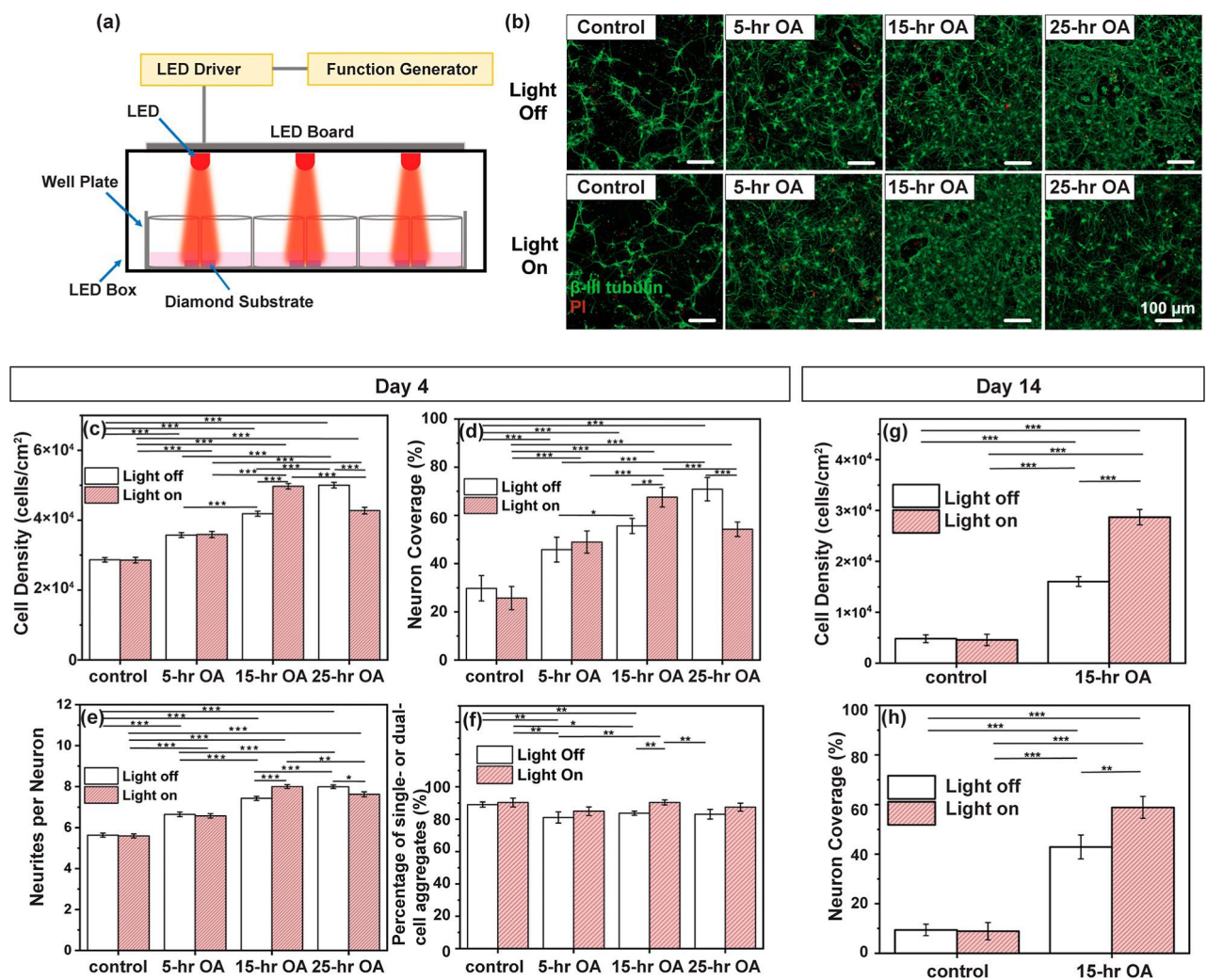

Figure 3. Immunohistochemistry of cells on diamond and control with or without light exposure. (a) Diagram of the light exposure setup comprising a PCB board equipped with six 810-nm LED bulbs, an LED box, an LED driver, and a function generator. A 24-well culture place is shown positioned within the LED box. (b) Representative fluorescent microscope images of neurons at Day 4 on the coverslip controls, as well as the 5-, 15-, and 25-h OA N-UNCD electrodes with or without NIR stimulation. Neurons were identified by *β*-III tubulin staining (green), while nuclei are visualized through propidium iodide (PI) staining (red). Quantitative analysis of neuron structures on Day 4 (c−f) and Day 14 (g,h). The graphs show (c,g) cell density (cell/cm$^2$), (d,h) neuron coverage (quantified as the area percentage represented by *β*-III tubulin staining), (e) the number of neurites per neuron, and (f) the percentage of single-cell or dual-cell aggregates on the coverslip controls and the OA N-UNCD surfaces under conditions with or without NIR exposure. Error bars represent standard errors. Statistical significance is denoted by \**p* < 0.05, \*\**p* < 0.01, \*\*\**p* < 0.001.

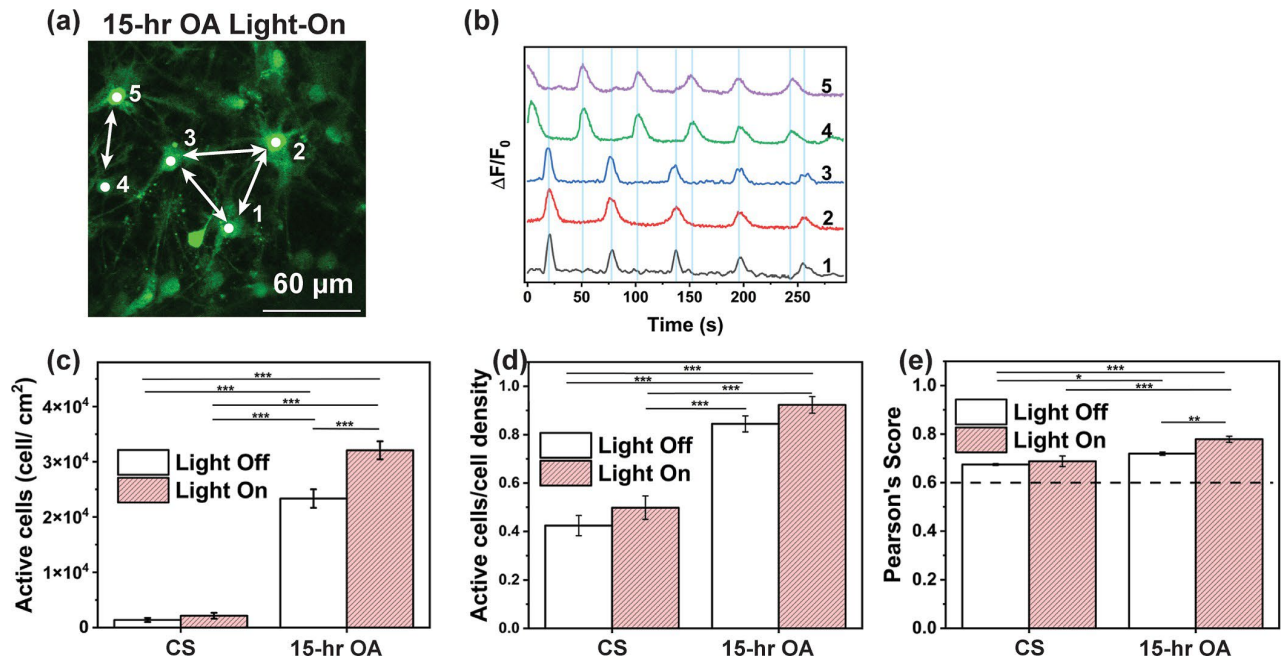

Figure 4. Intracellular calcium activity in neurons on Day 14. (a). Representative images of neurons loaded with Fluo-4 AM on a 15-h OA N-UNCD substrate after NIR stimulation on Day 14. The corresponding calcium transients ($\Delta F/F_0$ vs time) of the neurons labelled 1−5 are shown in (b). Connections indicated by arrows were determined via Pearson's cross-correlation analysis. (c) Density of cells exhibiting spontaneous calcium activity. (d) Ratio of spontaneously active cells relative to the total cell count. (e). Pearson's cross-correlation analysis was applied to data collected under various conditions, where a score of 0.6 (dotted line) denotes a high level of cross-correlation. Statistical significance is denoted by *$p < 0.05$, **$p < 0.01$, ***$p < 0.001$.

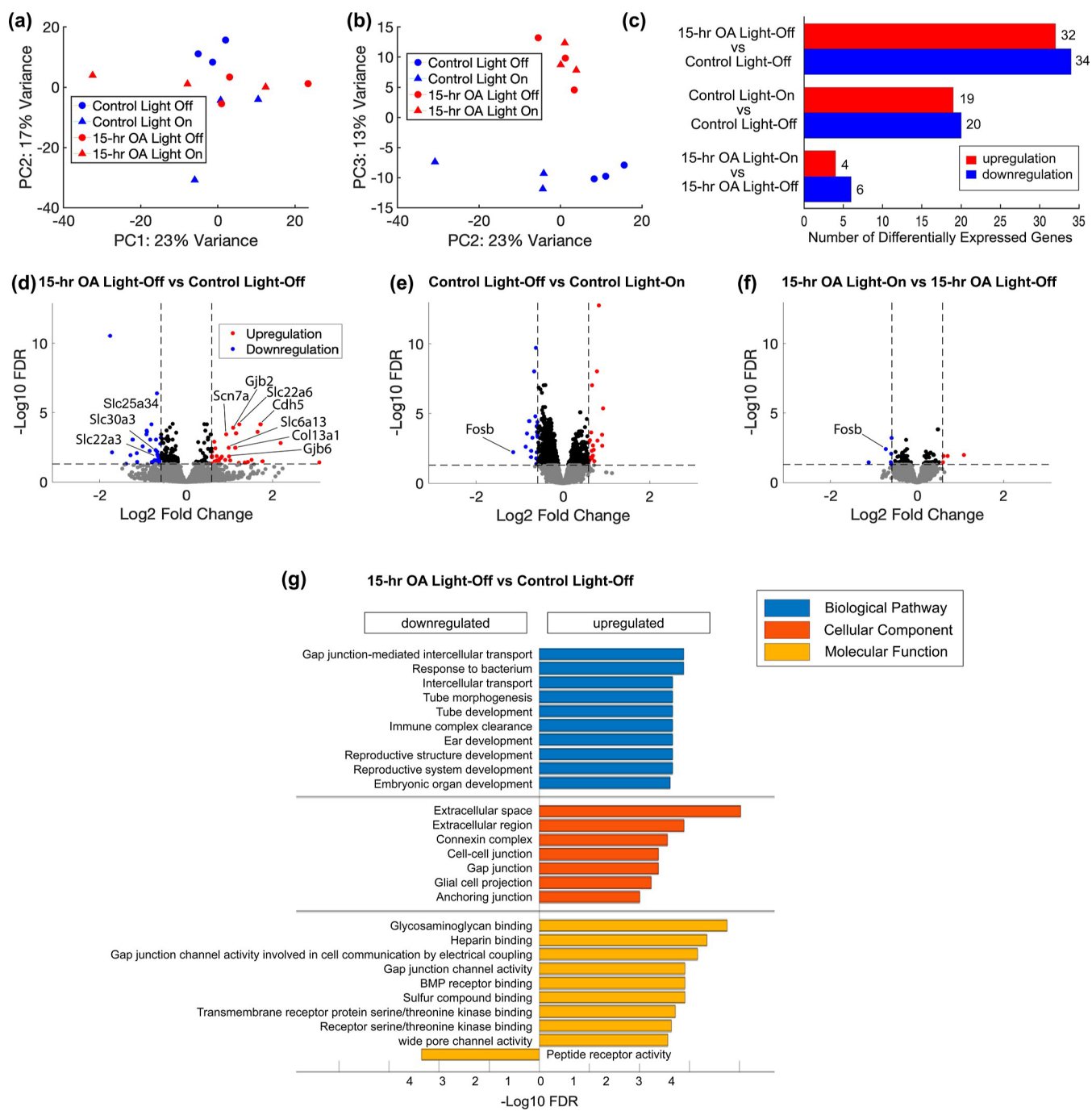

Figure 5. Comparative analysis of transcriptional profiles in neurons on day 14. (a,b). Principal component analysis highlights the variation in gene expression across different conditions, where PC3 exhibits a strong correlation with substrate type ($p = 1.39 \times 10^{-6}$). (c). A comprehensive summary of differentially expressed genes among different conditions. (d−f) Volcano plot illustrating gene expression, with dotted lines representing the cutoff threshold for analysis ($p = 0.05$, Fold Change = 1.5). (g) Gene Ontology (GO) enrichment analysis: the bar charts display upregulated (right) and downregulated (left) GO terms in the respective conditions